\documentclass[onecolumn,leqno]{amsproc}%
\usepackage{amsfonts}
\usepackage{amsmath}
\usepackage{amssymb}
\usepackage{graphicx}
\usepackage[none]{hyphenat}%
\theoremstyle{plain}

\numberwithin{equation}{section}
\allowdisplaybreaks[3]
\begin{document}
\title{}

\begin{center}
{\LARGE \textbf{A new market model in the large volatility case}}

\bigskip

\bigskip

\textbf{YUKIO HIRASHITA}

\bigskip

\bigskip

\textbf{Abstract}
\end{center}

\noindent We will compare three types of prices, namely, rational (hedging)
prices, geometric (growth rate) prices, and martingale (measure) prices. We
will show that rational prices in the complete market theory are sometimes
contrary to common sense. In the continuous-time case, we insist that the
market model should differ between the small volatility case $(\sigma
^{2}/2\leq r)$ and the large volatility case $(r<\sigma^{2}/2)$.

\bigskip



\bigskip

\begin{center}
{\large \textbf{1. Rational prices}}

\bigskip
\end{center}

\noindent Consider a complete market (single-step Cox-Ross-Rubinstein) model
in which

\ \ \ \ \ the riskless asset is $B_{0}\rightarrow B_{1}=B_{0}(1+r)$,

\ \ \ \ \ the risky asset is $\ \ \ S_{0}\rightarrow S_{1}=\left\{
\begin{array}
[c]{ccc}%
S_{0}(1+b) & \cdots & p\\
S_{0}(1+a) & \cdots & q=1-p
\end{array}
\right.  $,

\noindent where $-1<a<r<b$ with probability $P\{a\}=q>0$ and $P\{b\}=p>0$ (see
Shiryaev [9], page 408). In the complete market theory, the values of five
parameters $\{r,$ $a,$ $b,$ $p,$ $S_{0}\}$ can be independently provided under
the above conditions. We will insist that $S_{0}$ must be determined by $\{r,$
$a,$ $b,$ $p\}.$

We consider the contingent claim $f=S_{1}$. The rational price of $f$ is
$S_{0}$ because of the hedging portfolio $(0,$ $1)$. Here, the portfolio
$(\alpha,$ $\beta)$ implies investment $\alpha B_{0}+\beta S_{0}$. Moreover,
the rational price of the contingent claim $g=S_{0}(1+b)$ is $S_{0}%
(1+b)/(1+r)$ because of the hedging portfolio $(S_{0}(1+b)/(B_{0}(1+r)),$ $0)$.

It is easy to see that if $q=1/n$, then $E(|f-g|^{2})$ $=S_{0}^{2}(b-a)^{2}/n$
and $\lim_{n\rightarrow\infty}f=g$ (a.e.).

For example, if $B_{0}=1,$ $S_{0}=1,$ $a=0.1$ $<$ $r=0.2$ $<$ $b=11,$ and
$q=0.01,$ then
\begin{align*}
&  \text{the rational price of }f=\left\{
\begin{array}
[c]{ccc}%
12 & \cdots & p=0.99\\
1.1 & \cdots & q=0.01
\end{array}
\right.  \text{ is }1,\\
&  \text{the rational price of }g=\left\{
\begin{array}
[c]{ccc}%
12 & \cdots & p=0.99\\
12 & \cdots & q=0.01
\end{array}
\right.  \text{ is }10.
\end{align*}
The coexistence of these prices is contrary to common sense. Moreover,
\[
\text{the rational price of }\left\{
\begin{array}
[c]{ccc}%
108 & \cdots & p=0.99\\
-1 & \cdots & q=0.01
\end{array}
\right.  \text{ is 0,}%
\]
because of the hedging portfolio $(-10,$ $10)$. This is again contrary to
common sense. It is worth noting that a contingent claim is not necessarily
nonnegative, as is shown in \O ksendal [5], Section 12.2.

\newpage

\begin{center}
{\large \textbf{2. Geometric prices}}

\bigskip
\end{center}

\noindent We introduce the geometric price of a contingent as follows.

\bigskip

\textbf{Definition 2.1. }The geometric price $u$ of a contingent claim with
return $h(x)$ and distribution $F(x)$ is given by the equation%
\[
\sup_{_{\substack{0\leq z\leq1\\z\leq\text{ess }\inf_{x}\text{ }h(x)\text{
}z/u+1}}}\exp\left(  \int\log\left(  h(x)\text{ }z/u-z+1\right)  dF(x)\right)
=1+r,
\]
under certain auxiliary conditions (see Hirashita [3]).

\bigskip

\textbf{Theorem 2.2. }\textit{Suppose} $E:=\alpha p+\beta q>0,$ $0\leq p\leq1$
\textit{and} $r>0,$ \textit{then the geometric price }$u$\textit{ of
}$\left\{
\begin{array}
[c]{ccc}%
\alpha & \cdots & p\\
\beta & \cdots & q=1-p
\end{array}
\right.  $ \textit{is }$u=\alpha^{p}\beta^{q}/(1+r)$ (\textit{the discounted
price of the geometric mean})\textit{ and the optimal proportion of
investment} $z=1$, \textit{if }$\alpha>0,$ $\beta>0$ \textit{and} $\alpha
^{p}\beta^{q}/(1+r)$ $\leq1/(p/\alpha+q/\beta).$ \textit{Otherwise},\textit{
}$u$\textit{ and }$z$ \textit{are determined from the system of equations}%
\[
\left\{
\begin{array}
[c]{c}%
\left(  \frac{E-\alpha}{u-\alpha}\right)  ^{q}\left(  \frac{E-\beta}{u-\beta
}\right)  ^{p}=1+r,\\
z=\frac{(E-u)u}{(\alpha-u)(u-\beta)}.
\end{array}
\right.
\]
\textit{However, if }$\alpha\beta<0$\textit{ and }$(1-E/\alpha)^{q}%
(1-E/\beta)^{p}\leq1+r$\textit{, then, as }$r$\textit{ is too large, the above
equations have no solution.}

Proof. In the case where the optimal proportion of investment $z<1$ exists,
Definition 2.1 reduces to the system%
\[
\left\{
\begin{array}
[c]{c}%
\left(  \frac{\alpha z}{u}-z+1\right)  ^{p}\left(  \frac{\beta z}%
{u}-z+1\right)  ^{q}=1+r,\\
p\left(  \frac{\alpha}{u}-1\right)  \left(  \frac{\beta z}{u}-z+1\right)
+q\left(  \frac{\beta}{u}-1\right)  \left(  \frac{\alpha z}{u}-z+1\right)
=0,\\
0<\frac{\alpha z}{u}-z+1,\\
0<\frac{\beta z}{u}-z+1,
\end{array}
\right.
\]
which leads to the conclusion.\textit{
\ \ \ \ \ \ \ \ \ \ \ \ \ \ \ \ \ \ \ \ \ \ \ \ \ \ \ \ \ \ \ \ \ \ \ \ \ \ \ \ \ \ \ \ \ \ \ \ \ \ \ \ \ \ \ \ \ \ \ \ \ }%
$\square$

\bigskip

\textbf{Corollary 2.3.} \textit{Consider the special case of Theorem 2.2,
where }$p=q=1/2$ \textit{and} $\alpha>\beta$ $\geq0$. \textit{If} $E$
$\leq(1+r)\sqrt{\alpha\beta},$\textit{ then }$u=\sqrt{\alpha\beta}/(1+r)$
\textit{and }$z=1$\textit{. Otherwise, }$u=\kappa\alpha+(1-\kappa)\beta$
\textit{and} $z=$ $(E$ $-u)$ $u/((\alpha-u)$ $(u-\beta))$, \textit{where
}$\kappa$\textit{ }$=(1-\sqrt{1-1/(1+r)^{2}})/2$\textit{.}

\bigskip

For example, if $B_{0}=1,$ $S_{0}=1,$ $a=0.1$ $<$ $r=0.2$ $<$ $b=11,$ and
$q=0.01$, then%

\[
\text{the geometric price of }f=\left\{
\begin{array}
[c]{ccc}%
12 & \cdots & p=0.99\\
1.1 & \cdots & q=0.01
\end{array}
\right.  \text{ is }9.764\text{,}%
\]
because of%
\[
\frac{12^{0.99}1.1^{0.01}}{1.2}\doteq9.764<\frac{1}{\frac{0.99}{12}%
+\frac{0.01}{1.1}}\doteq10.918.
\]
Moreover, we obtain that%
\[
\text{the geometric price of }\left\{
\begin{array}
[c]{ccc}%
108 & \cdots & p=0.99\\
-1 & \cdots & q=0.01
\end{array}
\right.  \text{ is }86.079
\]
from the equation%
\[
\left(  \frac{106.91-108}{u-108}\right)  ^{0.01}\left(  \frac{106.91+1}%
{u+1}\right)  ^{0.99}=1.2,
\]
where $E=108\times0.99+(-1)\times0.01=106.91$ and $(1-106.91/108)^{0.01}%
(1+106.91)^{0.99}$ $\doteq98.349>1.2.$

\bigskip

\begin{center}
{\large \textbf{3. Martingale prices}}

\bigskip
\end{center}

\noindent The martingale measure (which is independent of the original
probability $p$) of the risky asset $S_{1}$ is given by%
\[
S_{1}^{\ast}=\left\{
\begin{array}
[c]{ccc}%
S_{0}(1+b) & \cdots & p^{\ast}=(r-a)/(b-a),\\
S_{0}(1+a) & \cdots & q^{\ast}=(b-r)/(b-a).
\end{array}
\right.
\]
\textit{ }The martingale price $(\alpha p+\beta q)/(1+r)$ of $\left\{
\begin{array}
[c]{ccc}%
\alpha & \cdots & p\\
\beta & \cdots & q=1-p
\end{array}
\right.  $ is obtained based on the assumption that the original measure is
the martingale measure when $\alpha\neq\beta$, that is, $\alpha=S_{0}^{\prime
}(1+b^{\prime}),$ $\beta=S_{0}^{\prime}(1+a^{\prime}),$ $p$ $=(r-a^{\prime})$
$/(b^{\prime}-a^{\prime})$, and $q=(b^{\prime}-r)/(b^{\prime}-a^{\prime}).$

For example, if $r=0.2$, then
\[
\text{the martingale price of }\left\{
\begin{array}
[c]{ccc}%
12 & \cdots & p=0.99\\
1.1 & \cdots & q=0.01
\end{array}
\right.  \text{ is }9.909\text{,}%
\]
which is similar to the geometric price of $9.764$.

As the martingale price $(\alpha p+\beta q)/(1+r)$, which is the discounted
price of the expectation with respect to the original measure, is independent
of the variance, the following contingent claims have the same martingale
price of $50$.
\[
\left\{
\begin{array}
[c]{ccc}%
60 & \cdots & p=0.5\\
60 & \cdots & q=0.5
\end{array}
\right.  ,\text{ }\left\{
\begin{array}
[c]{ccc}%
119 & \cdots & p=0.5\\
1 & \cdots & q=0.5
\end{array}
\right.  ,\text{ }\left\{
\begin{array}
[c]{ccc}%
120 & \cdots & p=0.5\\
0 & \cdots & q=0.5
\end{array}
\right.  ,\text{ }\left\{
\begin{array}
[c]{ccc}%
160 & \cdots & p=0.5\\
-40 & \cdots & q=0.5
\end{array}
\right.  .
\]
Most investors will pay $50$ for the claim $\left\{
\begin{array}
[c]{ccc}%
60 & \cdots & p=0.5\\
60 & \cdots & q=0.5
\end{array}
\right.  ,$ however, many investors will not pay $50$ for the claim $\left\{
\begin{array}
[c]{ccc}%
119 & \cdots & p=0.5\\
1 & \cdots & q=0.5
\end{array}
\right.  $. These tendencies need not be explained by the risk-aversion
mind-set, because the geometric prices of the above four contingent claims are
$50,$ $27.387,$ $26.834$, and $4.723$, respectively.

\bigskip

\begin{center}
{\large \textbf{4. Discussion}}

\bigskip
\end{center}

\noindent The history of the debate between the growth rate criterion and
expected utility is found in Christensen [2]. Samuelson [8] insists that
\textquotedblleft Pascal will always put all his wealth into the risky
gamble\textquotedblright\ $\left\{
\begin{array}
[c]{ccc}%
2.7 & \cdots & p=0.5\\
0.3 & \cdots & q=0.5
\end{array}
\right.  $ with price $1,$ \textquotedblleft according to the max $EX_{T}$
criterion.\textquotedblright\ With the given price $u=1$, the growth rate
function (see Definition 2.1)
\[
\exp\left(  \int\log\left(  h(x)\text{ }z/u-z+1\right)  dF(x)\right)  =\left(
\frac{2.7z}{1}-z+1\right)  ^{0.5}\left(  \frac{0.3z}{1}-z+1\right)  ^{0.5}%
\]
attains its maximum $12/\sqrt{119}\doteq1.100$ at the proportion of investment
$z=50/119\doteq0.420$. Therefore, we insist that Pascal will always put $42\%$
of his wealth into the risky gamble $\left\{
\begin{array}
[c]{ccc}%
2.7 & \cdots & p=0.5\\
0.3 & \cdots & q=0.5
\end{array}
\right.  $ with price $1.$ Therefore, we suspect Samuelson's assertion that
\textquotedblleft To maximize the geometric mean, one must stick only to
cash.\textquotedblright

\bigskip

\begin{center}
{\large \textbf{5. In the continuous-time case}}

\bigskip
\end{center}

\noindent The Black-Merton-Scholes model is given by

\ \ \ \ \ the riskless asset is $B_{t}=B_{0}e^{rt}$ \ $(t\geq0)$,

\ \ \ \ \ the risky asset is $\ \ \ S_{t}=S_{0}e^{(\mu-\sigma^{2}/2)t+\sigma
W_{t}}$ \ $(t\geq0)$,

\noindent where $W=(W_{t})_{t\geq0}$ is a Brownian motion (see Shiryaev [9],
page 739).

\bigskip

(1) If the original measure is the martingale measure, then $\mu=r$ (see
Shiryaev [9], page 765) and $(S_{t}/B_{t})_{t\geq0}$ is a martingale. In this
case, we have $S_{t}=S_{0}e^{(r-\sigma^{2}/2)t+\sigma W_{t}}$, $E(S_{t}%
)=S_{0}e^{rt}$, and $V(S_{t})/E(S_{t})^{2}=e^{\sigma^{2}t}-1$. Let
$G_{t}=e^{-\sigma^{2}t/2+\sigma W_{t}}$, then $(G_{t})_{t\geq0}$ is a
martingale with $E(G_{t})=1$ and $V(G_{t})=e^{\sigma^{2}t}-1$. The martingale
price of the riskless asset $S_{0}e^{rt}$ is $S_{0}$, and the martingale price
of the risky asset $S_{t}=S_{0}e^{rt}\times G_{t}$ is also $S_{0}$,
irrespective of the size of volatility $\sigma$. This is contrary to common sense.

\bigskip

(2) If $(\log(S_{t}/B_{t}))_{t\geq0}$ is a martingale, then $\mu=r+\sigma
^{2}/2$ and vice versa. In this case, we have
\[
S_{t}=S_{0}e^{rt+\sigma W_{t}},
\]
$E(S_{t})=S_{0}e^{(r+\sigma^{2}/2)t}$, and $V(S_{t})/E(S_{t})^{2}%
=e^{\sigma^{2}t}-1$. The condition that the optimal proportion of investment
is equal to $1$ is given by%
\begin{align*}
&  \int\frac{1}{h(x)}dF(x)\text{ \ }\exp\left(  \int\log h(x)dF(x)\right)  \\
&  =\frac{1}{\sqrt{2\pi t}}\int_{-\infty}^{\infty}\frac{1}{S_{0}e^{rt+\sigma
x}}e^{-x^{2}/(2t)}dx\text{ }\exp\left(  \frac{1}{\sqrt{2\pi t}}\int_{-\infty
}^{\infty}\left(  \log S_{0}+rt+\sigma x\right)  e^{-x^{2}/(2t)}dx\right)  \\
&  \leq e^{rt}\text{,}%
\end{align*}
which is equivalent to $\sigma^{2}/2\leq r$ (see Hirashita [3], Lemma 4.17,
Corollary 5.3, and Section 6). Therefore, the assumptions $\mu=r+\sigma^{2}/2$
and $\sigma^{2}/2\leq r$ (the small volatility case) deduce that the geometric
price
\[
\frac{\exp\left(  \int\log h(x)dF(x)\right)  }{e^{r}}=\frac{\exp\left(
\frac{1}{\sqrt{2\pi t}}\int_{-\infty}^{\infty}\left(  \log S_{0}+rt+\sigma
x\right)  e^{-x^{2}/(2t)}dx\right)  }{e^{rt}}=S_{0}%
\]
of $S_{t}$ at the start time $0$ is independent of $t$.

\bigskip

(3) We consider a market model where $r<\sigma^{2}/2$ (the large volatility
case). For example, if $r=0.04$ and $\sigma=0.4$, then the geometric prices of%
\[
S_{t}=S_{0}e^{(r-0.00616)t+\sigma W_{t}}%
\]
$(0\leq t\leq2)$ at the start time $0$ are approximately constant $S_{0}$ or,
more specifically, included in the interval $[S_{0},$ $1.00033S_{0})$. This
can be shown by applying the two-dimensional Newton-Raphson method to the
system of equations (cf. Definition 2.1, $1+r\rightarrow e^{rt}$)%
\[
\left\{
\begin{array}
[c]{c}%
\frac{1}{\sqrt{2\pi t}}\int_{-\infty}^{\infty}\log\left(
e^{(r-0.00616)t+\sigma x}z/u-z+1\right)  e^{-x^{2}/(2t)}dx=rt,\\
\int_{-\infty}^{\infty}\frac{e^{(r-0.00616)t+\sigma x}-u}%
{e^{(r-0.00616)t+\sigma x}z-uz+u}e^{-x^{2}/(2t)}dx=0.
\end{array}
\right.
\]

It is worth noting that the volatility of stocks is typically in the interval
$0.2\leq\sigma\leq0.5$ (Hull [5], page 238). For example, if $\sigma\leq0.5$
and $r\geq0.01$, then there exits $c=c(\sigma$, $r)\geq0$ such that the
geometric prices of $S_{t}=S_{0}e^{(r-c)t+\sigma W_{t}}$ $(0\leq t\leq1)$ at
the start time $0$ are included in the interval $[S_{0},$ $1.0052S_{0})$.

\bigskip

\begin{center}
\textbf{References}

\bigskip
\end{center}

\noindent\lbrack1] F. Black and M. Scholes, The pricing of options and
corporate liabilities,

J. Political Economy 81 (1973), 637-654.

\noindent\lbrack2] M. M. Christensen, On the History of the Growth Optimal
Portfolio. University

of Southern Denmark, Working Paper, 2005.

\noindent\lbrack3] Y. Hirashita, Game pricing and double sequence of random variables,

Preprint (2007), arXiv:math.OC/0703076.

\noindent\lbrack4] Y. Hirashita, Delta hedging without the Black-Scholes
formula, Far East

J. Appl. Math. 28 (2007), 157-165.

\noindent\lbrack5] J. Hull, Options, futures, and other derivatives, Prentice
Hall, New Jersey,

2003.

\noindent\lbrack6] D. G. Luenberger, Investment science, Oxford University
Press, Oxford, 1998.

\noindent\lbrack7] B. \O ksendal, Stochastic differential equations, Springer,
Berlin, 1998.

\noindent\lbrack8] P. A. Samuelson, The \textquotedblleft
Fallacy\textquotedblright\ of Maximizing the Geometric Mean in Long

Sequences of Investing or Gambling. Proc. Nat. Acad. Sci. USA. 68 (1971),

2493-2496.

\noindent\lbrack9] A. N. Shiryaev, Essentials of stochastic finance, World
Scientific, Singapore,

1999

\bigskip

\bigskip

\noindent Faculty of Liberal Art

\noindent Chukyo University

\noindent Yagoto-honmachi 101-2, Showaku

\noindent Nagoya, Aichi 466-8666, Japan

\noindent e-mail: yukioh@cnc.chukyo-u.ac.jp

\end{document}